\documentclass[prl,twocolumn,floatfix]{revtex4-1}
\usepackage{graphicx,bm,color}
\usepackage{times}
\usepackage{amsmath,amssymb}
\usepackage{epstopdf}

\begin{document}


\title{Ultrafast X-ray Absorption Spectroscopy of Strongly Correlated Systems: Core Hole Effect}

\author{Chen-Yen Lai}
\author{Jian-Xin Zhu}
\affiliation{Theoretical Division, Los Alamos National Laboratory, Los Alamos, New Mexico 87545, USA}
\affiliation{Center for Integrated Nanotechnologies, Los Alamos National Laboratory, Los Alamos, New Mexico 87545, USA}

\date{\today}

\begin{abstract}
In recent years, ultrafast pump-probe spectroscopy has provided insightful information about nonequilibrium dynamics of excitations in materials.
In a typical experiment of time-resolved x-ray absorption spectroscopy, the systems are excited by a femtosecond laser pulse (pump pulse) followed by an x-ray (probe pulse) after a time delay to measure the absorption spectra of the photoexcited systems.
We present a theory for nonequilibrium x-ray absorption spectroscopy in one-dimensional strongly correlated systems.
The core hole created by x-ray is modeled as an additional effective potential of the core hole site which changes the spectrum qualitatively.
In equilibrium, the spectrum reveals the charge gap at half-filling and the metal-insulator transition in the presence of the core hole effect.
Furthermore, a pump-probe scheme is introduced to drive the system out of equilibrium before the x-ray probe.
The effects of the pump pulse with varying frequencies, shapes and fluences are discussed for the dynamics of strongly correlated systems in and out of resonance.
The spectrum indicates that the driven insulating state has a metallic droplet around the core hole.
The rich structures of the nonequilibrium x-ray absorption spectrum give more insight into the dynamics of electronic structures.
\end{abstract}

\maketitle

The primary goal of x-ray spectroscopy is to probe the properties of core level electrons and their coupling to the electrons near Fermi energy~\cite{Ament:2011jy,Olovsson:2013cf}.
Contrary to the angle-resolved photoemission spectroscopy, which provides an accurate measurement of low-energy band structure~\cite{Damascelli:2003kq,Zhu:2006fg,Kordyuk:2014jt,Avella:2015wr}, x-ray spectroscopy offers a sensitive and versatile probe of the high-energy excitations.
On the other hand, the rapidly developed resonant inelastic x-ray scattering, which is a photon-in photon-out process, provides more information on the excitation spectrum~\cite{Ament:2011ez,Dean:2016ee,Huang:2016fi,Minola:2015dn}.
The short time evolution of the slightly excited initial state in both bosonic and fermionic systems can be exploited to answer fundamental questions in condensed matter physics and strongly correlated systems.
In cold atom systems where the atoms have relatively slow motions~\cite{Chien:2015kc},
several studies have investigated the relaxation of the quantum state after sudden quench~\cite{Lai:2016hr,Lai:2017kq} and
the proposed scheme to probe the properties of the many-body state~\cite{Senaratne:2018he,Bohrdt:2018gz,Stewart:2008kt}.
Ultrafast laser spectroscopy provides an additional gear to investigate the electronic structure of excited states in materials.
Although the photoexcited carriers usually have short lifetimes~\cite{MiajaAvila:2016cu}, the state-of-the-art pump-probe technique can still study the time evolution of the materials,
for instance, cuprate superconductors~\cite{Fausti:2011dy,Huang:2016fi,Matsuzaki:2015dn}, transition metal oxides~\cite{Sheu:2013kn,Gandolfi:2017ds}, and charge density wave compounds~\cite{Hellmann:2012dn,Yamakawa:2017go,Gomi:2014gj,Matsuzaki:2015dn}.
Along with the development of a tabletop x-ray source~\cite{Weisshaupt:2014he,Popmintchev:2018hf}, the reconstruction of the charge, spin, and lattice dynamics~\cite{Berner:2013kp,Guarise:2014kw} from time-resolved x-ray spectroscopy is within reach.
The obtained insight will be very helpful in understanding emergent phenomena in strongly correlated electron systems (among which, the Mott insulator-metal transition is one intriguing phenomenon and the properties of excited state spectrum are difficult to measure).
By using x-ray absorption in experiments~\cite{Bruno:2014gh,Torriss:2017kz,Preziosi:2018kr}, one can determine the metal-insulator transition as the temperature varies or the doping changes~\cite{Fausti:2011dy}.
The dynamics of such systems, driven out of equilibrium by external stimuli, can provide insight into the underlying interactions between different coupling mechanisms within femto- to picosecond timescales~\cite{Giannetti:2016hp,Ligges:2018br}.
Selective measurement techniques are necessary to probe specific excitations because
the connection between various types of excitations is hidden deep in the quantum wave function, which cannot be observed directly.

Different from other probe techniques, x-ray absorption spectroscopy (XAS) also brings out the core hole effect~\cite{Weijs:1990kk,Mauchamp:2009ia,Suzuki:2012fv,Fuggle:1980bu}.
Combined with the valance-electron quantum dynamics, the core hole effect is expected to create novel phenomena in nonequilibrium systems.
In this Letter, we propose a single band model to study both the static and nonequilibrium (NE)-XAS of one-dimensional strongly correlated systems.
We model the core hole created by the incident x-ray as an attractive potential for the valence electrons~\cite{si}.
In equilibrium, the spectrum reveals the metal-insulator transition for systems at the half-filling due to the core hole effect.
The NE spectra have even more features, including a metallic droplet around the core hole from a driven insulating state.
Furthermore, the NE-XAS shows a resonance between the frequency of the incident pump pulse and the charge gap of the systems.

{\it Theoretical formalism. --}
Starting from a conventional two-orbital model~\cite{si} and considering the dipole matrix element: $\langle 3d_\sigma \vert T_\sigma \vert 2p \rangle$ between two orbitals
for absorption where $T_\sigma$ is a dipole transition operator, we propose an effective single band model to capture the x-ray absorption spectrum.
In equilibrium, the valance electrons are described by a Fermi-Hubbard model (FHM),
\begin{equation}\label{eq:fhm}
    \mathcal{H}=-J\sum_{\langle ij \rangle, \sigma}(d^\dagger_{i\sigma}d_{j\sigma} + H.c.) + U\sum_in_{i\uparrow}n_{i\downarrow},
\end{equation}
where $d^\dagger_{i\sigma}$ is fermion creation operator with spin $\sigma$ at site $i$ and the density operator is $n_{i\sigma}\!=\!d^\dagger_{i\sigma}d_{i\sigma}$.
Hereafter, the hopping amplitude is set to unity $J\!=\!1$ and the time unit is $t_0\!=\!\hbar/J$.
In general, the XAS can be determined from the Fermi golden rule
\begin{equation}
    \mathcal{I}_{\text{XAS}}(\omega) = \sum_\sigma\sum_F \vert\langle F \vert d_{m\sigma}^\dagger \vert I \rangle\vert^2 \delta(E_F-E_I-\hbar\omega).
\end{equation}
Here, $\vert I\rangle$ ($\vert F\rangle$) and $E_{I(F)}$ are the initial (final) states and energies, and $d_{m\sigma}^\dagger$ denotes the electron excited from the core level to the valence band with spin $\sigma$ at site-$m$.
Using the identity
\begin{equation}
  \delta(x)\!=\!-\frac{1}{\pi}\lim_{\Gamma\rightarrow0^+}\text{Im}\left\{\frac{1}{x+i\Gamma}\right\},
\end{equation}
the intensity can be expressed as
\begin{equation}
  \mathcal{I}_{\text{XAS}}(\omega)\!=\! -\frac{1}{\pi}\sum_\sigma\text{Im}\mathcal{A}_\sigma(\omega),
\end{equation}
with the quantity $\mathcal{A}_\sigma(\omega)$ given by
\begin{eqnarray}\label{eq:SpectralOmega}
    \mathcal{A}_\sigma(\omega) &=& -i\int_0^\infty dt e^{i\omega t}e^{-\Gamma t}\mathcal{A}_\sigma(t)\\
    &=& -i\int_0^\infty dt e^{i\omega t}e^{-\Gamma t} \!\langle I \vert e^{i\mathcal{H}t} d_{m\sigma} e^{-i\mathcal{H}_mt} d_{m\sigma}^\dagger \vert I \rangle.\nonumber
\end{eqnarray}
Here, the $\Gamma$ represents the core hole lifetime broadening effect.
$\mathcal{H}_m$ inside the nonlocal time correlation function $\mathcal{A}(t)$ is the sum of the equilibrium Hamiltonian $\mathcal{H}$ and the effective attractive potential $-V_\text{ch}\sum_\sigma n_{m\sigma}$ due to the presence of the core hole.
In some of the transition metal oxides, the typical core-valence interaction is about $30\%$ stronger than the valence-valence interaction~\cite{si,hariki2017}.
The initial state, $\vert I \rangle$, is the many-body wave function right before the x-ray probe kicks in.
Below we consider two situations: (i) the initial state as the ground state of the Hamiltonian and (ii) a NE initial state encoding the effect of the pump pulse.

{\it Static XAS. --}
In the equilibrium case, we use the ground state of Eq.~\eqref{eq:fhm} as the initial state, which can be obtained by density matrix renormalization group accurately~\cite{White:1992ie,Schollwock:2011gl,Lai:2017kq}.
For noninteracting fermions, the wave function is a Slater determinant, which can be expressed as a matrix product state (MPS)~\cite{PietroSilvi:2012eh,Schollwock:2011gl}.
Here, different initial interactions and filling fractions ($\bar{n}_f\!=\!\sum_{i\sigma}n_{i\sigma}/L$ where $L$ is the number of lattice sites) are studied.
The nonlocal time correlation function is solved in a time evolving block decimation~\cite{Vidal:2003ug,Vidal:2004jc,Feiguin:2005fk} under the MPS framework~\cite{Daley:2004hk,McCulloch:2007gia,Lai:2008ez,Lai:2016kh,Lai:2017kq}.
Because the initial state is the ground state of the FHM, the evolution operator acting on the bra state can be reduced to a phase factor, i.e., $\langle \text{GS} \vert e^{i\mathcal{H}t} \!=\! e^{iE_\text{GS}t}\langle \text{GS} \vert$ with ground state energy $E_\text{GS}$.
Due to the finite lifetime of the core hole, the simulation of real time dynamics is not required to be long to capture the spectrum quantitatively.
Throughout this work, we use a time step of $\delta t\!=\!10^{-3}t_0$ in the second order Suzuki-Trotter approximation for time evolution and a core hole lifetime of $\Gamma\!=\!0.2J$ for calculations of the spectra.

\begin{figure}[t]
    \begin{center}
        \includegraphics[width=\columnwidth]{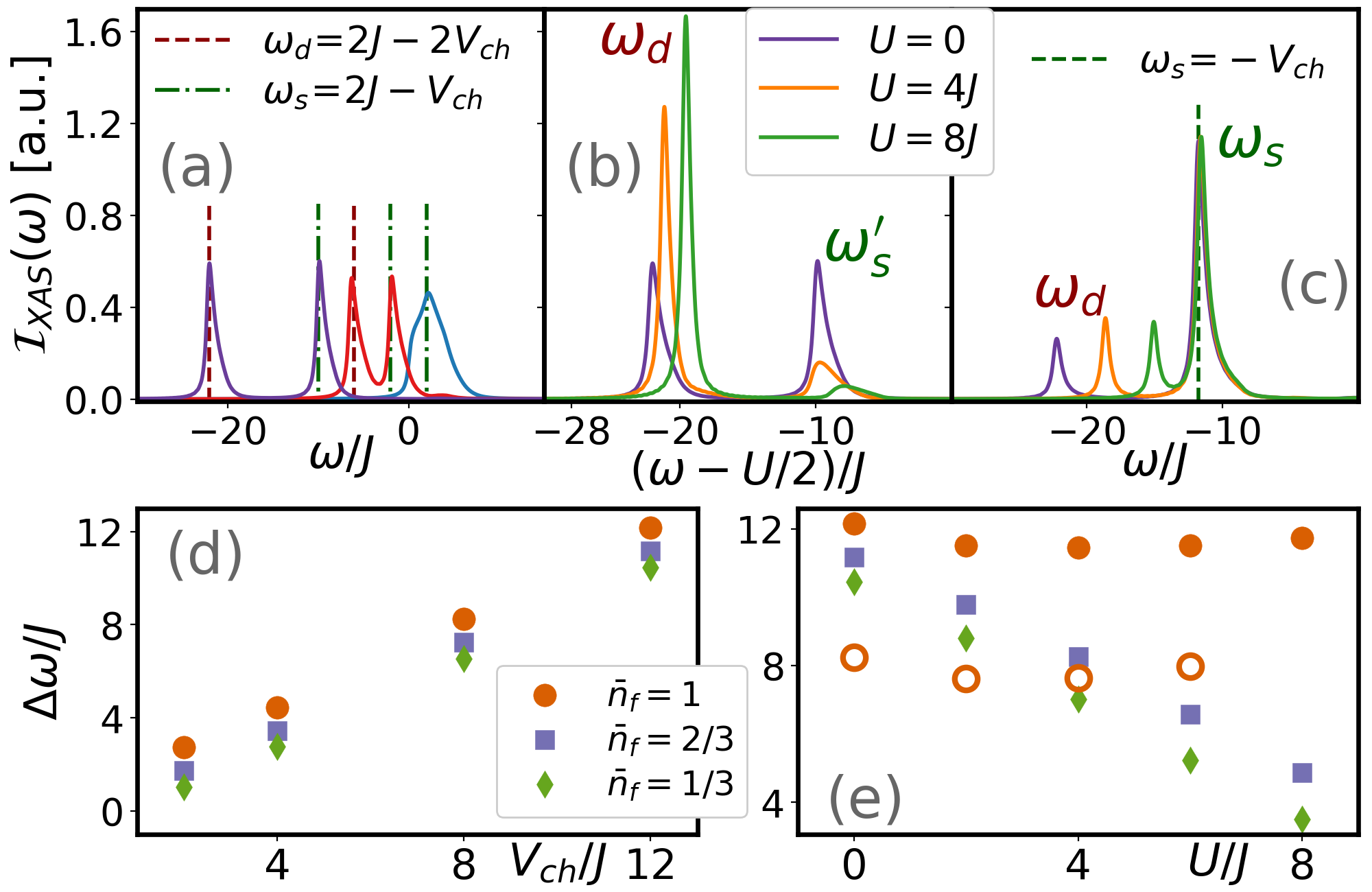}
        \caption{
            Static XAS under filling fraction (a)--(b) $\bar{n}_f\!=\!1$, and (c) $\bar{n}_f\!=\!2/3$.
            In Fig.~\ref{fig:U0Vch12}(a), the initial state is the Fermi sea state ($U\!=\!0$); and the different core hole potentials of $V_{ch}/J\!=\!0$ (blue), $4$ (red), and $12$ (purple) are considered.
            The spectrum splits in the presence of nonzero core hole potential.
            In Fig.~\ref{fig:U0Vch12}(b) and Fig.~\ref{fig:U0Vch12}(c), with a shared legend, the initial state is the ground state with different $U$ and the core hole potential is set to $V_\text{ch}\!=\!12J$.
            (d) Frequency difference versus core hole potential from Fermi sea state with different fillings.
            (e) Frequency difference versus interaction with different fillings, where the core hole potentials are set to $V_\text{ch}\!=\!12J$ (filled) and $V_\text{ch}\!=\!8J$ (open).
        }
        \label{fig:U0Vch12}
    \end{center}
\end{figure}

The main objective is to capture the core hole effect in the XAS of strongly correlated systems.
Starting from a noninteracting Fermi sea state at half-filling, which is shown in Fig.~\ref{fig:U0Vch12}(a), the results show that the spectrum is split into two peaks from one due to the core hole potential.
It is worth mentioning that the spectrum corresponds to the absorption part of the spectral density in the absence of core hole potential~\cite{si}.
The locations of the peak indicate the corresponding bound state energy due to the core hole potential, which is around $\omega_d$ ($\omega_s$) for a doubly (singly) occupied bound state at the core hole site, as marked in Fig.~\ref{fig:U0Vch12}(a), where the amplitudes of both peaks are roughly the same.
The difference of these two frequencies, $\Delta\omega\!=\!\omega_s\!-\!\omega_d$, reveals the core hole potential as shown in Fig.~\ref{fig:U0Vch12}(d) for all three different fillings.
For interacting fermions away from the half-filling, as shown in Fig.~\ref{fig:U0Vch12}(c), the singly occupied state still has stronger amplitude and all peaks are around $-V_\text{ch}$, which signals the presence of the core hole potential.
On the other hand, the doubly occupied state shifts in frequency as the interaction changes and is around $U\!-\!2V_\text{ch}$.
This explains the energy difference of $\Delta\omega\!=\!V_\text{ch}\!-\!U$, as shown in Fig.~\ref{fig:U0Vch12}(e).
Therefore, the XAS can enable us to determine the core hole potential and the interacting strength of the measured strongly correlated systems.
For systems at half-filling, three different interaction strength are compared in Fig.~\ref{fig:U0Vch12}(b).
In order to make a comparison to free fermions, the frequency is shifted to match the symmetric point determined by the density of states~\cite{si}.
Compared to the systems away from half-filling, where only the doubly occupied state has the frequency shifted by $U$, both peaks are now shifted due to the strongly correlated effects.
After the core electron is excited, the core hole site is nearly a doubly-occupied in this Mott insulating phase.
Because the filling factor is exactly at half originally, the excess electron forms a doublon even if the electron escapes from the core hole site.
This doublon outside the core hole site makes the frequency of the singly occupied bound state shift by $U$, as well as the doubly occupied bound state.
Therefore, in the half-filling, the energy of the singly occupied state is shifted to $\omega^\prime_s\!=\!U\!-\!V_\text{ch}$ and the frequency difference is independent of the interaction as two different core hole potentials, shown in Fig.~\ref{fig:U0Vch12}(e).

\begin{figure}[b]
    \begin{center}
        \includegraphics[width=\columnwidth]{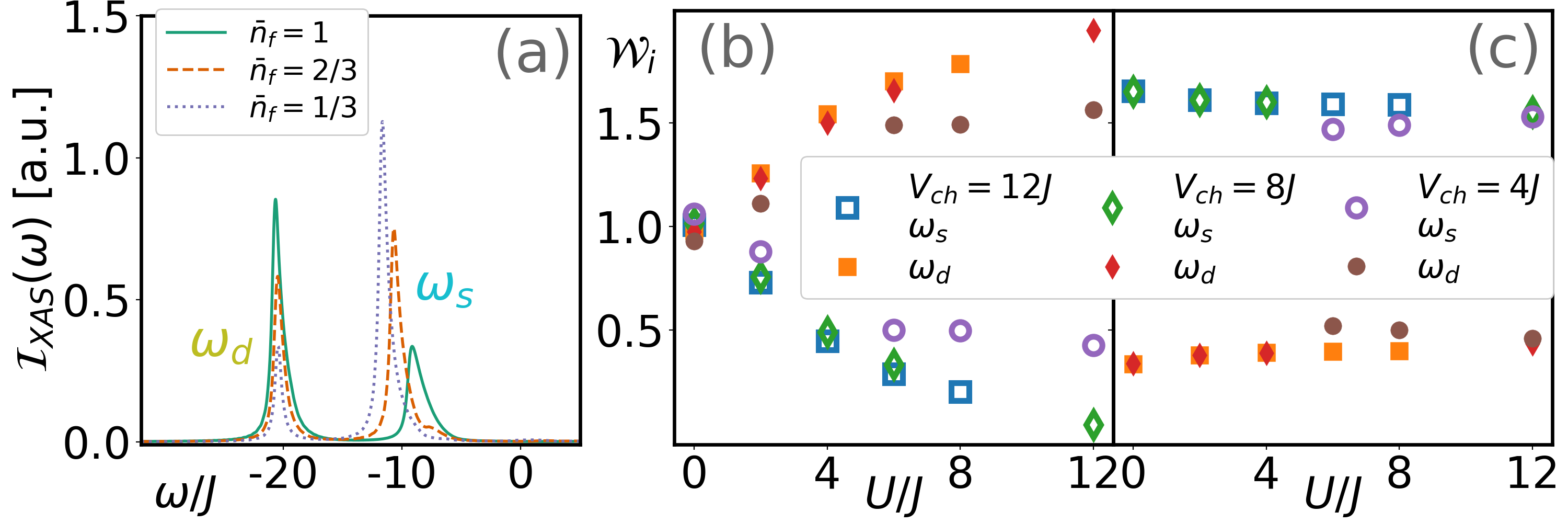}
        \caption{
            (a) XAS for different fillings under the same interaction ($U\!=\!2J$) and core hole potential ($V_\text{ch}\!=\!12J$).
            Weights of XAS from singly (open symbols) and doubly (filled symbols) occupied states: (b) $\bar{n}_f\!=\!1$, and (c) $\bar{n}_f\!=\!1/3$ under different interactions and core hole potentials.
        }
        \label{fig:EqmWeight}
    \end{center}
\end{figure}

In addition, the weight of the corresponding response reveals important information about the electronic structure.
It is defined as
\begin{equation}
  \mathcal{W}_i=\int^{\omega_i+\delta\omega}_{\omega_i-\delta\omega}\mathcal{I}_\text{XAS}(\omega)d\omega,
\end{equation}
where $\delta\omega$ is a finite width that covers the decay tail due to broadening.
The spectrum is normalized such that $\sum_i\mathcal{W}_i\!\approx\!2$ due to the spin degrees of freedom.
It is known that the charge gap exists in the one-dimensional half-filled FHM with any finite $U$ in the thermodynamic limit~\cite{Lieb:1968ic,Lieb:2003hl}.
From Fig.~\ref{fig:U0Vch12}(b), we can immediately observe this feature.
In the absence of interaction, both singly and doubly occupied bound states have almost the same weight.
As the interaction increases, the weight of the doubly occupied state always dominates over the singly occupied state.
We notice that the chemical potential shift is not only manifesting in the change of the dominant peak position, but also in the relative intensity transfer between $\omega_d$ and $\omega_s$ peaks~\cite{vanaken2002,tan2012}.
Also, this weight transfer depends on the strengths of both the Hubbard interaction and the core hole potential.
Figures~\ref{fig:EqmWeight}(b) and~\ref{fig:EqmWeight}(c) give a more quantitative analysis on the shifting weight for varying interaction strengths and core hole potentials.
At the half-filling, both peaks have equal weight from the Fermi sea state, and the signals from the doubly occupied state become more dominant as the interaction becomes finite for various core hole potentials,
especially in a deep core hole potential [compare the weights for $V_\text{ch}\!=\!4J$ and $8J$ in Fig.~\ref{fig:EqmWeight}(b)].
This suggests that the charge gap opens in finite interactions.
In the low electron occupation limit, the dominant signal is always the singly occupied state as shown in Fig.~\ref{fig:EqmWeight}(c).
In the weak interaction regime of $U\!<\!4J$, the weight distributions are almost identical despite different core hole potentials.

\begin{figure}[t]
    \begin{center}
        \includegraphics[width=\columnwidth]{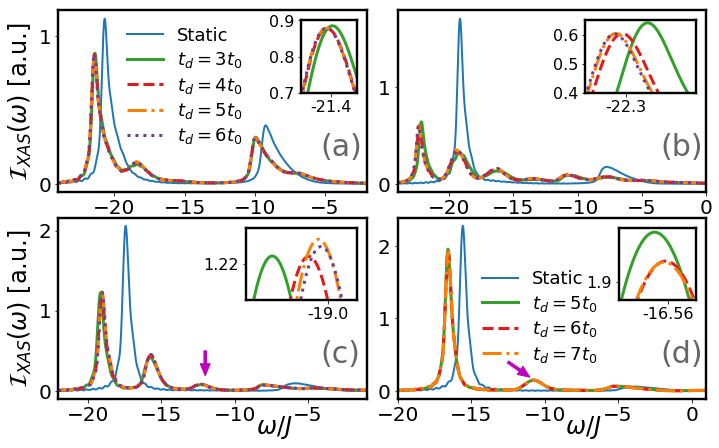}
        \caption{
            NE-XAS with core hole potential $V_\text{ch}\!=\!12J$ and different time delays for (a) $U\!=\!2J$, (b) $4J$, (c) $6J$, and (d) $8J$ under pulse fluences of $\Omega t_0\!=\!3$, $\tau\!=\!2t_0$, and $A_0\!=\!0.1$ in Fig.~\ref{fig:TRHalf}(a)--~\ref{fig:TRHalf}(c), and $A_0\!=\!0.3$ in Fig.~\ref{fig:TRHalf}(d).
            In Fig.~\ref{fig:TRHalf}(c) and~\ref{fig:TRHalf}(d), the peak around $\omega\!\sim\!-12J$ (magenta arrow) emerges as a singly occupied core hole signal from the metallic droplet.
            The system is at the half-filling.
            }
        \label{fig:TRHalf}
    \end{center}
\end{figure}

{\it Nonequilibrium XAS. --}
When a laser pulse is incident before the x-ray photon, the initial state in Eq.~\eqref{eq:SpectralOmega} is no longer the ground state of the FHM.
We model the effect of the laser pulse via a time-dependent Pieles phase in the Hamiltonian of $J\!\rightarrow\!Je^{iA(t)}$, where the phase has a Gaussian profile:
\begin{equation}
  A(t)\!=\!A_0e^{-(t+t_d)^2/2\tau^2}\cos\Omega(t+t_d),
\end{equation}
with the intensity as $A_0$, the central frequency as $\Omega$, the pulse shape with width as $\tau$, and the time delay $t_d$.
(We set the probe as always starting at $t=0$.)
The average incoming number of photons per lattice site from the pump is estimated to be $\propto\!A_0^2\Omega\tau$.
In order to capture the effect from the pump pulse, the time delay is chosen to be large enough where the amplitude of the pulse has almost vanished [$A(0)\!<\!0.1$] before measuring the XAS.
Therefore, the real time dynamics of the initial ground state wave function under the pulse needs to be simulated before calculating the nonlocal time correlation function.
In other words, the initial state in Eq.~\eqref{eq:SpectralOmega} is given by $\vert I \rangle\!=\!\hat{U}(-2t_d,0)\vert \text{GS} \rangle$, where $\hat{U}$ is the time evolution operator of the FHM, including the interaction with the electromagnetic field.

We first vary only the time delay by keeping the pulse intensity, the frequency, and the shape fixed.
The NE-XAS for the half-filling is shown in Fig.~\ref{fig:TRHalf} with different time delays $t_d$.
Because our model does not include the relaxation effect, the NE-XAS will never recover back to the equilibrium one even after an extended time delay.
Here, the spectra from different time delays do not change much because the state only picks up some extra phases after the tail of the pulse diminishes.
Small changes of the frequency and amplitude (see the insets of Fig.~\ref{fig:TRHalf}) are due to the infinitesimal tail of the pulse.
As long as the time delay is long enough, the signals become translational invariant in time as one compares $t_d\!=\!5t_0$ and $6t_0$.
Also, the shifting of the peaks is roughly equal to the energy changes of the state from the pumping.
Besides that, the spectrum is qualitatively different from the one at equilibrium.
First of all, there are only two major peaks in the equilibrium spectrum, but the NE-XAS exhibits much richer features from the excited states.
For the weak interaction case (e.g., $U\!=\!2J$), new peaks emerge around $\omega_{d(s)}+3J$, where the shift matches the pump pulse frequency $\Omega$.
As the interaction increases to $4J$, the fluence from the pulse is severe, and this is because the Mott gap is close to the frequency of the pump pulse.
This resonance effect will be elaborated on later.
Once the interaction reaches $6J$ and $8J$, a new peak emerges around $-V_\text{ch}$.
As we already mentioned, in the equilibrium spectrum, this energy corresponds to the singly occupied state ($\omega_s$) from a metallic state.
Because the photoemission spectrum shows that the system remains gapped~\cite{si}, the results conjecture that this metallic signal is induced by the core hole and should be a droplet around the core hole site.
A similar effect was reported in superconductors with an impurity or disorder~\cite{Balatsky:2006ce}.

\begin{figure}[t]
    \begin{center}
        \includegraphics[width=\columnwidth]{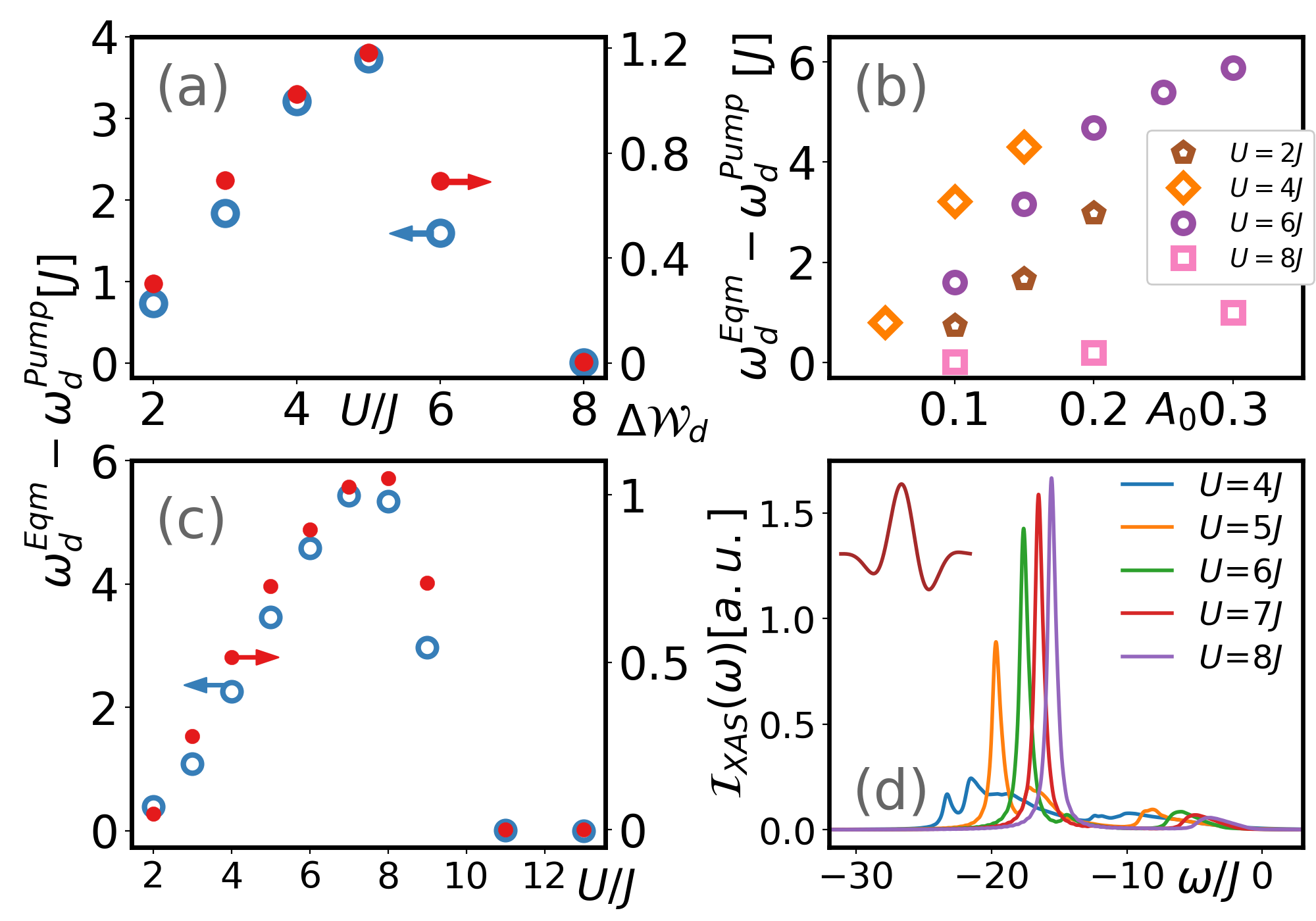}
        \caption{
            (a) Changes of weight (filled symbols) and frequency (open symbols) of the doubly occupied state versus interaction under the intensity of $A_0\!=\!0.1$.
            (b) Frequency shift of the doubly occupied state versus different intensities under different interactions.
            In Fig.~\ref{fig:TR60W3}(a) and~\ref{fig:TR60W3}(b), the pump pulses have $\Omega t_0\!=\!3$, $\tau\!=\!2t_0$, and $t_d\!=\!6t_0$.
            (c) Changes of weight (filled symbols) and frequency (open symbols) of the doubly occupied state versus interaction when systems at half-filling under fluence of $\Omega t_0\!=6\!$, width of $\tau\!=\!3t_0$, time delay of $t_d\!=\!6t_0$, and $A_0\!=\!0.1$.
            (d) NE-XAS for systems at half-filling with different $U$ under single cycle terahertz pulse fluence of $\Omega t_0\!=1\!$, width of $\tau\!=\!3t_0$, time delay of $t_d\!=\!6t_0$ and intensity of $A_0\!=\!1$.
            Inset shows the terahertz pulse profile.
            Core hole potential is set to $V_\text{ch}\!=\!12J$ in all panels.
        }
        \label{fig:TR60W3}
    \end{center}
\end{figure}

We then vary the intensity of the pulse to study the quantitative change of the NE-XAS by using the same time delay of $t_d\!=\!6t_0$ to ensure the pump pulse is almost finished.
In Fig.~\ref{fig:TR60W3}, a quantitative analysis of the nonequilibrium XAS us shown under different shapes of the pump pulse.
For all interaction strengths, both the singly and doubly occupied peaks get smaller weight as the intensity increases.
One expects that the system will melt down and the spectrum will become completely featureless as the state is excited into the continuum when the pump pulse is very strong.
Before that, the spectrum appears to have peaks separated by the energy close to the pump pulse frequency.
It is more interestingly that the pump pulse used here has a frequency of $\Omega t_0\!=\!3$ and the strongest influence on the $U\!=\!5J$ state.
Considering the same intensity ($A_0\!=\!0.1$ and $0.15$, for example), the shift of the frequency of the doubly occupied state is larger for $U\!=\!5J$ as shown in Fig.~\ref{fig:TR60W3}(b).
As the interaction increases to $6J$ and $8J$, the effects of the frequency shift and the weight of the doubly occupied state are also smaller than the one of $U\!=\!5J$, for which a detailed comparison is shown in Fig.~\ref{fig:TR60W3}(a).
By switching the frequency to $\Omega t_0\!=\!6$, the detailed spectrum is shown in the Supplemental Material~\cite{si} and the shifting of frequency is shown in Fig.~\ref{fig:TR60W3}(c).
The shift of the frequency is bigger as the interaction increases and reaches its maximum around $8J$.
The effect diminishes once the interaction becomes stronger.
From the results of two different frequencies,
both the shift of frequency and the change in the weight of the doubly occupied state give the same conclusion that there is a resonance between the pump pulse frequency and the charge gap in the systems.
On the other hand, for the single cycle terahertz pulse with a frequency of $\Omega t_0\!=\!1$, as shown in Fig.~\ref{fig:TR60W3}(d), the results show that the spectrum becomes featureless for weak and intermediate interactions (e.g., $U\!\leq\!4J$).
As the interaction becomes stronger, the NE-XAS is less affected by the pump pulse.
For instance, $U\!=\!8J$, the shift of the frequency and the change in weight are minimal when compared to the static XAS.

{\it Conclusion. --}
We have proposed a single band model to capture the core hole effect in the XAS, and we calculated the spectrum of a one-dimensional strongly correlated system.
The static XAS is able to distinguish the corresponding core hole potential and interaction strength of the strongly correlated materials.
Due to the strongly correlated effect,
the static XAS reveals the charge gap from the doubly occupied bound state when the system is half-filling with a finite interaction.
Furthermore, considering the pump pulse with different time delays, intensities and frequencies, the NE-XAS show that the driven system has a metallic droplet around the core hole,
which is a similar phenomenon to the impurity influence on the electronic states of superconductors.
Our results have uncovered that the static and nonequilibrium XASs can help to identify the excitations contributing to the spectrum and guide the future pump-probe experiments on strongly correlated materials, such as Sr$_2$CuO$_{3+\delta}$~\cite{kidd2008,keren1993,motoyama1996} or other cuprate compounds with Cu-O corner (or edge) sharing chains~\cite{motoyama1996,hase1993,kim2006}.
Those materials can be successfully synthesized and some of them can be doped away from the half-filling, on which the photoemission spectrum has also been measured~\cite{kidd2008}.

\acknowledgments{
We thank Jhi-Shih You and Marton Kan\'{a}sz-Nagy for fruitful discussion in the early stage of this work.
This work was carried out under the auspices of the U.S. Department of Energy (DOE)  National Nuclear Security Administration under Contract No. 89233218CNA000001. It was supported by the Center for Integrated Nanotechnologies, a DOE Office of Science User Facility, and in part by the LANL LDRD Program.
The numerical programs were built upon universal tensor library~\cite{Kao:2015gb} and the computational resource was provided by the LANL Institutional Computing Program.
}


\end{document}


\title{Supplemental Material for\\ Ultrafast x-ray absorption spectroscopy of strongly correlated systems: Core hole effect}

\author{Chen-Yen Lai}
\author{Jian-Xin Zhu}
\affiliation{Theoretical Division, Los Alamos National Laboratory, Los Alamos, New Mexico 87545, USA}
\affiliation{Center for Integrated Nanotechnologies, Los Alamos National Laboratory, Los Alamos, New Mexico 87545, USA}

\date{\today}

\maketitle

\begin{figure}[b]
    \begin{center}
        \includegraphics[width=\columnwidth]{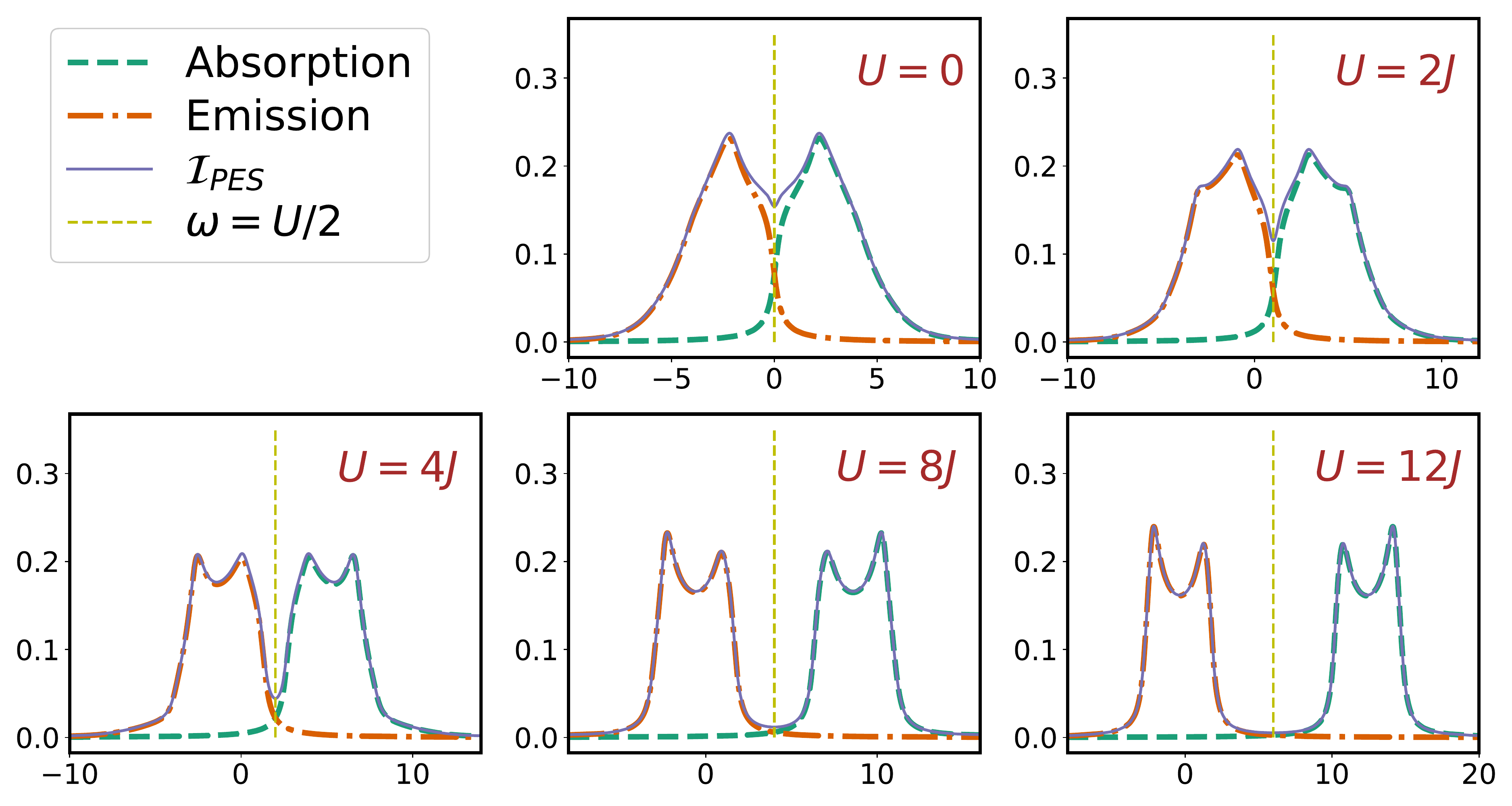}
        \caption{
            Complete photoemission spectrum $\mathcal{I}_\text{PES}(\omega)$ without core hole effect at half-filling under different interaction.
            For $U\!=\!0$, the two peaks are around $\pm2J$ due to the van Hove singularity.
            The symmetric position is at $U/2$ around upper and lower Hubbard band for interacting fermions.
            The broadening is set to $\Gamma\!=\!0.3J$.
        }
        \label{fig:HalfFillDOS}
    \end{center}
\end{figure}

\section{Effective single band model}
The minimum conventional model for XAS requires two orbitals, core ($p$) levels and valence ($d$) bands.
The valence band consist of itinerant electrons described by Fermi Hubbard model (FHM),
\begin{eqnarray}\label{eq:FHM}
    \mathcal{H}_{d} &=& -\sum_{\langle ij\rangle, \sigma}(Jd^\dagger_{i\sigma}d_{j\sigma} + h.c.) + U\sum_in^{(d)}_{i\uparrow}n^{(d)}_{i\downarrow}\;,
\end{eqnarray}
where the \(d^\dagger_{i\sigma}\) is creation operator of electrons in \(d\) orbital on site-$i$ with spin \(\sigma\), and the density operator is $n^{(d)}_{i\sigma}=d^\dagger_{i\sigma}d_{i\sigma}$.
The core levels have the Hamiltonian,
\begin{equation}
    \mathcal{H}_{p} = \sum_{i,\eta}\epsilon_p^{(\eta)}p^\dagger_{i\eta}p_{i\eta}\;,
\end{equation}
where \(p^\dagger_{i,\eta}\) is creation operator of core electron with quantum number \(\eta\!=\!(j,m_j)\) on site-$i$.
In the single photon excitation process, only one core electron is excited into the valance band.
We assume the zeroth order Slater integral (i.e. monopole part), $F^0_{pd}$, dominates in the $pd$ interaction.
In the density-density interaction approximation, the interactions between core electrons with different quantum number of $2p$-shell and the valance electron of $3d$-shell commute with each other, which enables us to treat the individual core hole excitation process independently.
Therefore, the coupling between an individual core level to be excited and valence state is given by
\begin{equation}
    \mathcal{H}_{dp} = \sum_i\tilde{V}_\text{ch}(n^{(p)}_i-1)n^{(d)}_i\;,
\end{equation}
where the number operators are $n_i^{(d)}\!=\!\sum_\sigma d^\dagger_{i\sigma}d_{i\sigma}$ and $n^{(p)}_{i}\!=\!p^\dagger_{i}p_{i}$\;.
Hereafter we have dropped the orbital index $\eta$.
This is the core-valence coupling expression commonly used~\cite{Nocera:2018,Benjamin:2014ek,Jia:2016,nagy:2016}.
Before the core electron is excited into the valance band, each core level is completely filled and the above Hamiltonian is irrelevant.

Here, we mainly focus on the Cu $L$-edge transition from $2p\!\rightarrow\!3d_{x^2-y^2}$ and follow the same procedure as in Ref.~\cite{Nocera:2018}.
From the Fermi-Golden rule, the absorption cross section is given by
\begin{equation}\label{eq:fermigolden}
    \mathcal{I}_\text{XAS}(\omega)\propto\sum_F\vert\langle \tilde{F} \vert \mathcal{\hat{D}} \vert \tilde{I} \rangle\vert^2 \delta(E_F-E_I-\hbar\omega)\;,
\end{equation}
where $\tilde{I} (\tilde{F})$ is the initial (final) state that has both core and valance electrons and the dipole transition operator is written as
\begin{equation}
    \mathcal{\hat{D}}\!=\!\sum_{m,\sigma}[A(\hat{\epsilon})d^\dagger_{m\sigma}p_{m}+h.c.]\;.
\end{equation}
Here, $A(\hat{\epsilon})\!=\!\langle 3d_{x^2-y^2} \vert \hat{\epsilon}\cdot\hat{r} \vert 2p\rangle$ is the matrix element of the dipole transition between the core $2p$ to the valance $3d_{x^2-y^2}$ and is set to unity.
We model the single x-ray photon event for the spectroscopy in the linear response regime and the core hole is completely localized due to the short lifetime.
Finally, we arrive at the single band model for the absorption cross section,
\begin{equation}
    \mathcal{I}_\text{XAS}(\omega)\propto\sum_{F,\sigma}\vert\langle F \vert d^\dagger_{m\sigma} \vert I \rangle\vert^2 \delta(E_F-E_I-\hbar\omega),
\end{equation}
which is Eq.~(2) in the main text and $I (F)$ is ths initial (final) state of valance electron only.
The core hole effect is coming from the core-valance coupling after the core hole is created.
Once the x-ray photon excites one core level electron to valence band which resulting in $n^{(p)}_m\!=\!0$, it creates an attractive potential $-\tilde{V}_\text{ch}$ to valence electrons at site-$m$.
The attractive potential is approximated as a core hole potential to the valence electron and we ignore the core levels after this,
\begin{equation}
    \mathcal{H}_{dp}\Rightarrow\mathcal{H}_\text{ch} = -V_\text{ch}n_m \;,
\end{equation}
where the superscript $(d)$ of the density operator is dropped.
It is worthy to mention that the effective value $V_\text{ch}$ here might be slightly different from the bare core-valence coupling $\tilde{V}_\text{ch}$ due to the screening effect.
On top of the effective impurity caused by the core hole, we need to create an electron on the site-$m$ of the valance band, $d^\dagger_m\vert I\rangle$ where $\vert I\rangle$ is the quantum many body wave function at the moment before the x-ray photon comes in.

\begin{figure}[b]
    \begin{center}
        \includegraphics[width=\columnwidth]{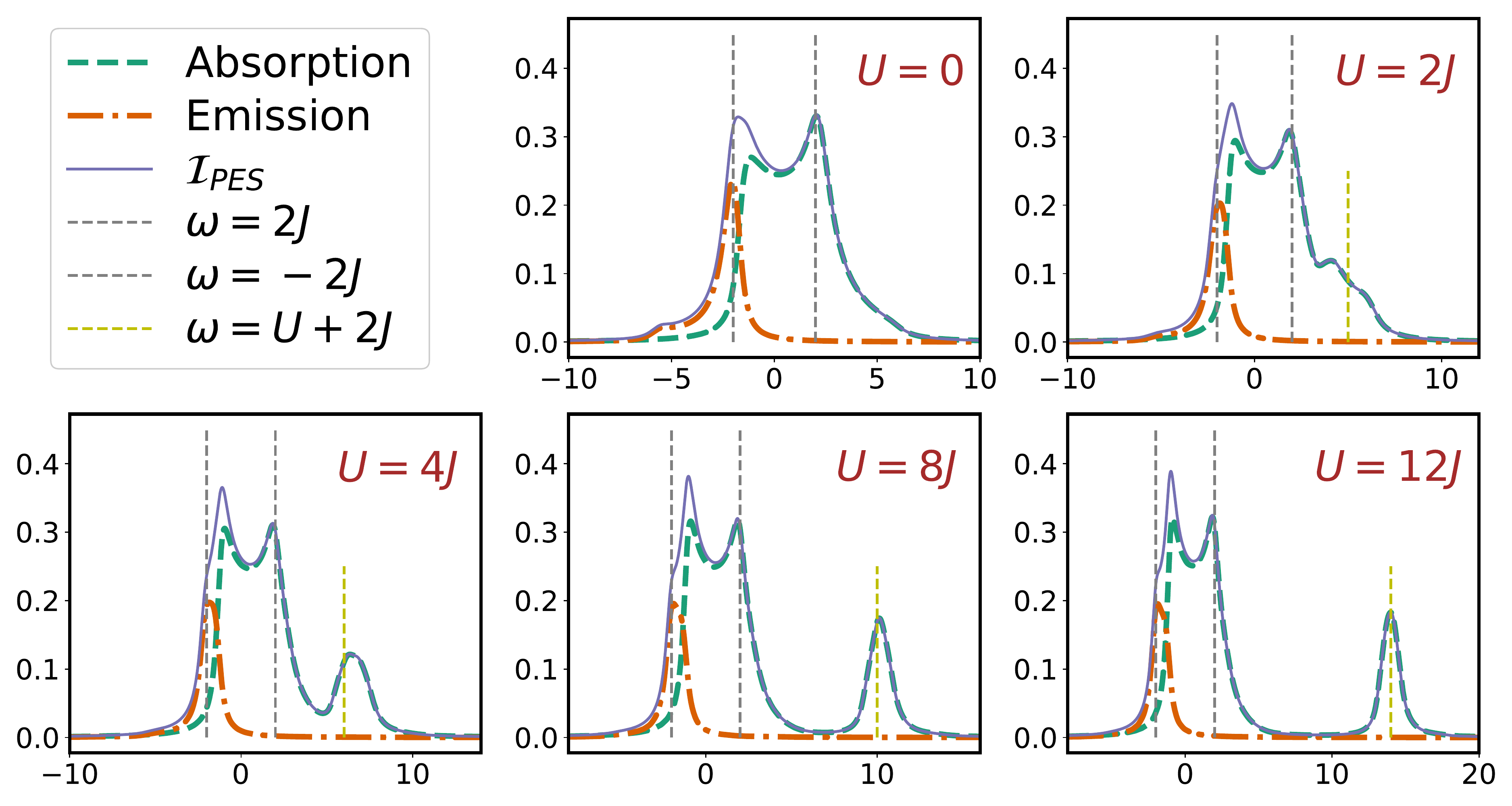}
        \caption{
            Complete photoemission spectrum $\mathcal{I}_\text{PES}(\omega)$ without core hole effect at one third filling under different interaction.
            The symmetric position is at $0$.
            The broadening is set to $\Gamma\!=\!0.3J$.
        }
        \label{fig:OneThirdDOS}
    \end{center}
\end{figure}
\section{Photoemission spectroscopy}
\subsection{Equilibrium}
Since our approach solve the ground state at canonical ensemble, the exact chemical potential is unknown.
To properly identify the exact Fermi energy to be our reference point in the spectrum, we calculate the density of states here.
In the absence of core hole potential $V_\text{ch}\!=\!0$, one will obtain half of the the density of states from photoemission spectroscopy.
By calculating the emission spectrum from
\begin{equation}
    \mathcal{I}_\sigma(\omega) = -i\int_0^\infty dt e^{i(\hbar\omega-E_I)t}e^{-\Gamma t}\langle I \vert d^\dagger_{m\sigma} U(-t,0) d_{m\sigma} \vert I \rangle,
\end{equation}
we can determine the zero energy reference point from the complete photoemission spectrum (PES), $\mathcal{I}_\text{PES}\!=\!\frac{1}{\pi}\sum_\sigma[\text{Im} \mathcal{A}_\sigma(\omega)+\text{Im}\mathcal{I}_\sigma(\omega)]$.
We demonstrate this calculation for half filling in Fig.~\ref{fig:HalfFillDOS} and one third filling in Fig.~\ref{fig:OneThirdDOS} for various interaction strength $U$.
For systems away from half filling and non-interacting fermions, the symmetry position is around $0$ which is the Fermi energy.
For insulating states, half filling with non zero $U$, the symmetry point is at $U/2$.

We remark that dynamical DMRG~\cite{Jeckelmann:2002dd,Kuhner:1999ef,Hallberg:2006jg} which use correction vector can determine density of state more accurately~\cite{Benthien:2004ee} than simulating in time domain or one needs to do linear prediction~\cite{Ganahl:2015ce} to improve the accuracy near $\omega\!=\!0$.
However, in the non-equilibrium, simulating in the time domain is more straight forward than the dynamical DMRG.

\subsection{Nonequilibrium}
\begin{figure}[t]
    \begin{center}
        \includegraphics[width=\columnwidth]{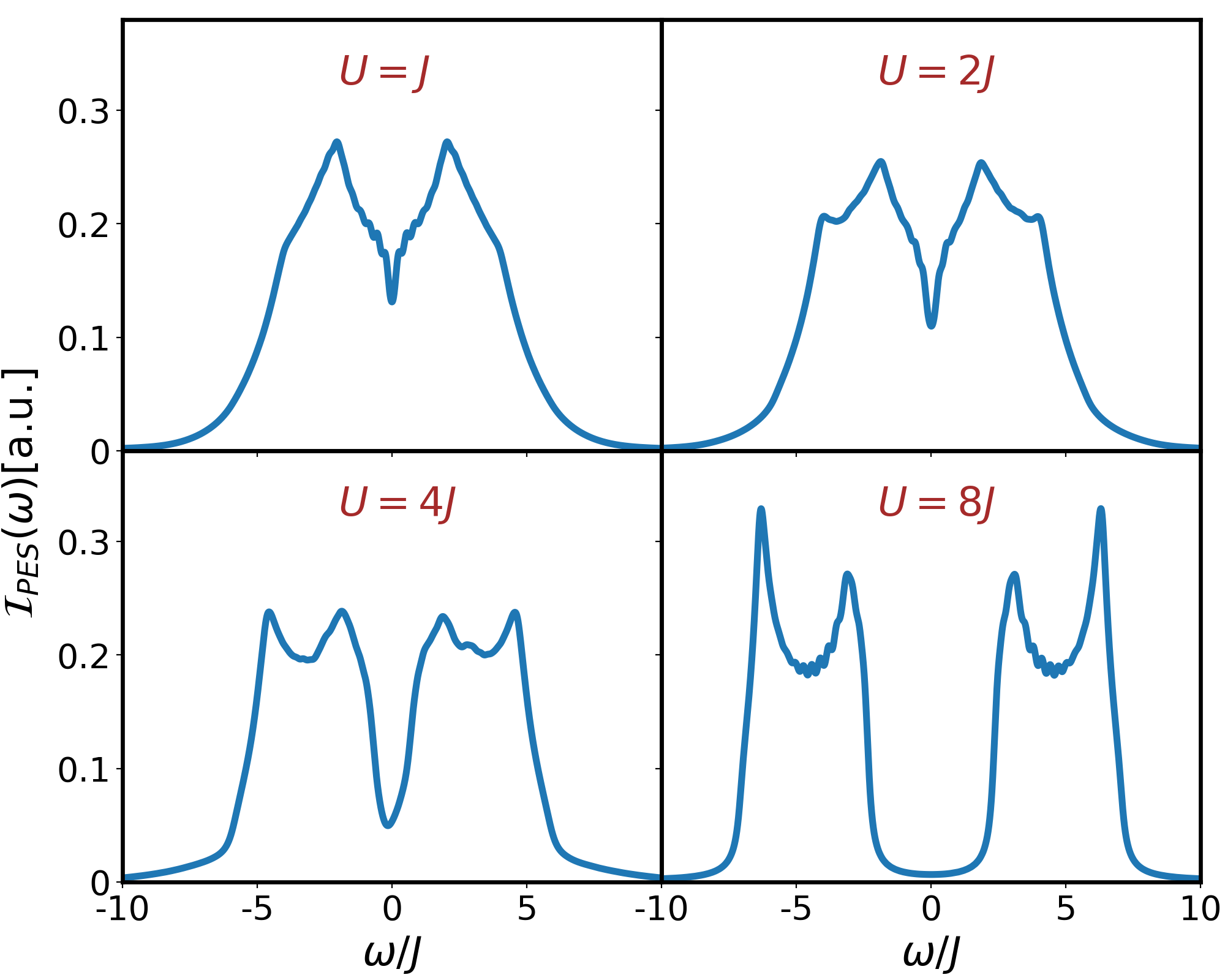}
        \caption{
            The photoemission spectrum of non-equilibrium state under the pulse influence $A_0\!=\!0.1$, $\Omega t_0\!=\!3$, $\tau\!=\!2t_0$, and time delay $t_d\!=\!6t_0$.
            The finite non-zero value in $U\!=\!8J$ is due to the broadening which is set to $\Gamma\!=\!0.15J$.
            The system is half-filled and all non-zero $U$ is a Mott insulator initially.
            The spectrum is shifted $U/2$ accordingly.
        }
        \label{fig:NE_DOS}
    \end{center}
\end{figure}

The system can be driven into a metallic state or remains an insulator after the pump pulse and we calculate the photoemission spectrum by using the many-body state under the pump pulse influence.
Figure~\ref{fig:NE_DOS} shows four different interactions under the same pump pulse.
It is clear that the gap vanishes after the initial insulating state undergoes the pump pulse fluence for small $U$.
For large and off-resonance pump pulse, like $U\!=\!8J$, the system remains gapped as a bulk insulating state.
In Fig.~3 of the main text, the emerged peak around energy $-V_\text{ch}$ signals that a metallic state might exist in the system.
We conjecture that this metallic droplet is induced by the dynamically emerged core hole from the x-ray probe since the photoemission spectrum show that the system remains insulator.
As one ramps up the intensity ($A_0$) of the pump pulse, we expect that the system will be driven to metallic state even for the large $U$.

\section{Nonequilibrium x-ray absorption spectroscopy}
\begin{figure}[b]
    \begin{center}
        \includegraphics[width=\columnwidth]{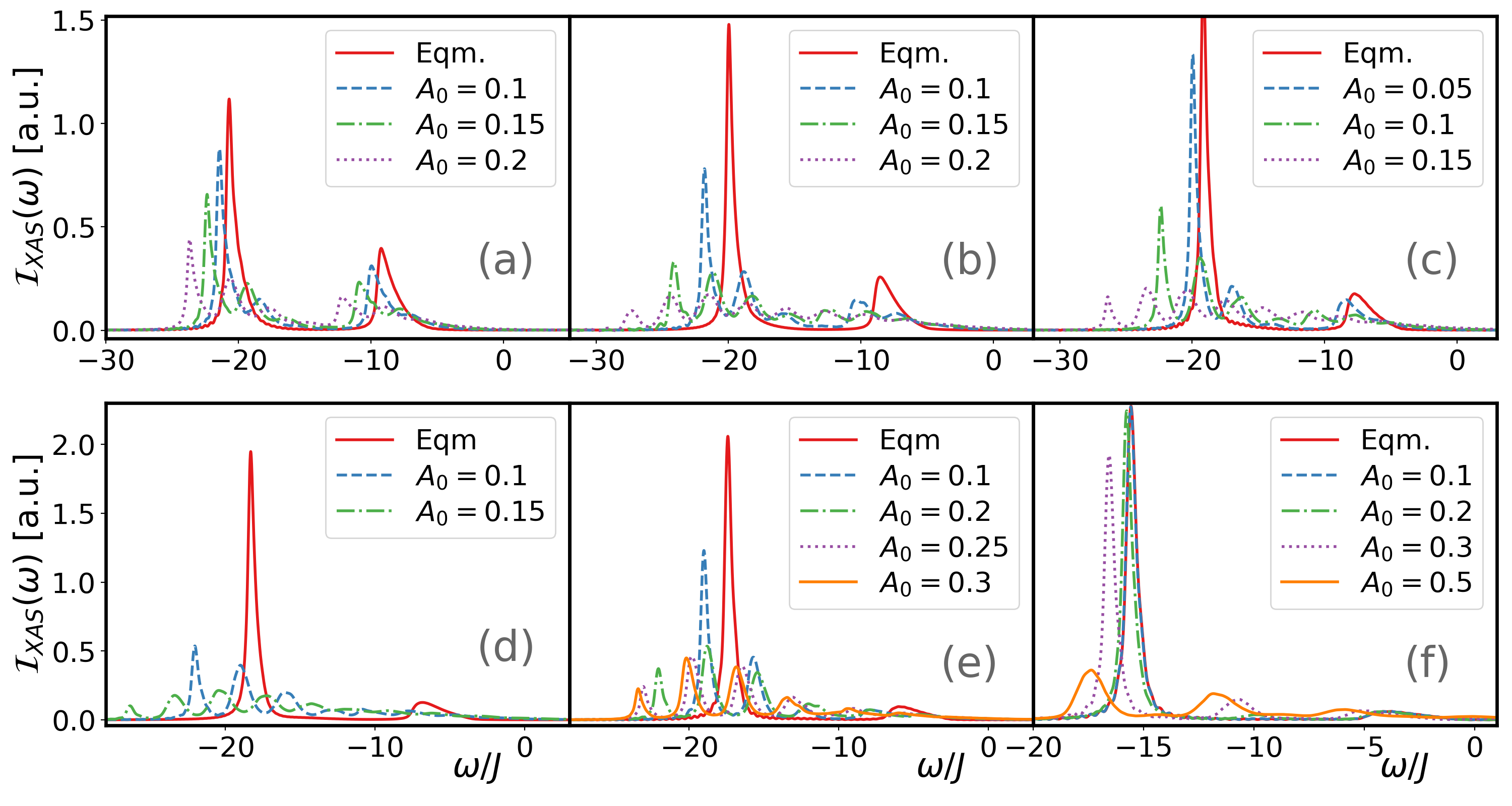}
        \caption{
            The NE-XAS for systems at half filling under different $A_0$'s where Eqm. stands for static XAS under different interactions (a) $U\!=\!2J$, (b) $U\!=\!3J$, (c) $U\!=\!4J$, (d) $U\!=\!5J$, (e) $U\!=\!6J$, (f) $U\!=\!8J$.
            The pump pulse is set to $\Omega t_0\!=\!3$, width $\tau\!=\!2t_0$, and time delay $t_d\!=\!6t_0$.
            The core hole potential is set to $V_\text{ch}\!=\!12J$.
        }
        \label{fig:AHalf}
    \end{center}
\end{figure}

Figure~\ref{fig:AHalf} shows the detail NE-XAS for pump pulse frequency $\Omega t_0\!=\!3$.
This further confirms our theory which states (i) the pump pulse causes the shifts of the frequency and change of the weight, (ii) new peaks emerges which reveals the pump pulse frequency, (iii) the resonance happens around $U\!\sim\!\Omega\!+\!2J\!=\!5J$ in this case, and finally (iv) the metallic signal caused by the dynamically emerged core hole in strong coupling.
From various different $U$'s, the frequency has blue shift and the weight of the major peak decreases as the intensity of the pump pulse increases.
It is clear that the pump pulse has strongest fluence on the $U\!=\!5J$ state where the system breaks down and the spectrum become featureless even the intensity of the pump pulse ($A_0$) is weak in Fig.~\ref{fig:AHalf}(d).
In Fig.~\ref{fig:AHalf}(f), the metallic signal emerges near $\omega\!\sim\!-12J$ when the intensity reaches $A_0\!=\!0.2$.
The peak is more visible when the intensity is $0.3$ as we show in the Fig.~3(d) of the main text.

\begin{figure}[t]
    \begin{center}
        \includegraphics[width=\columnwidth]{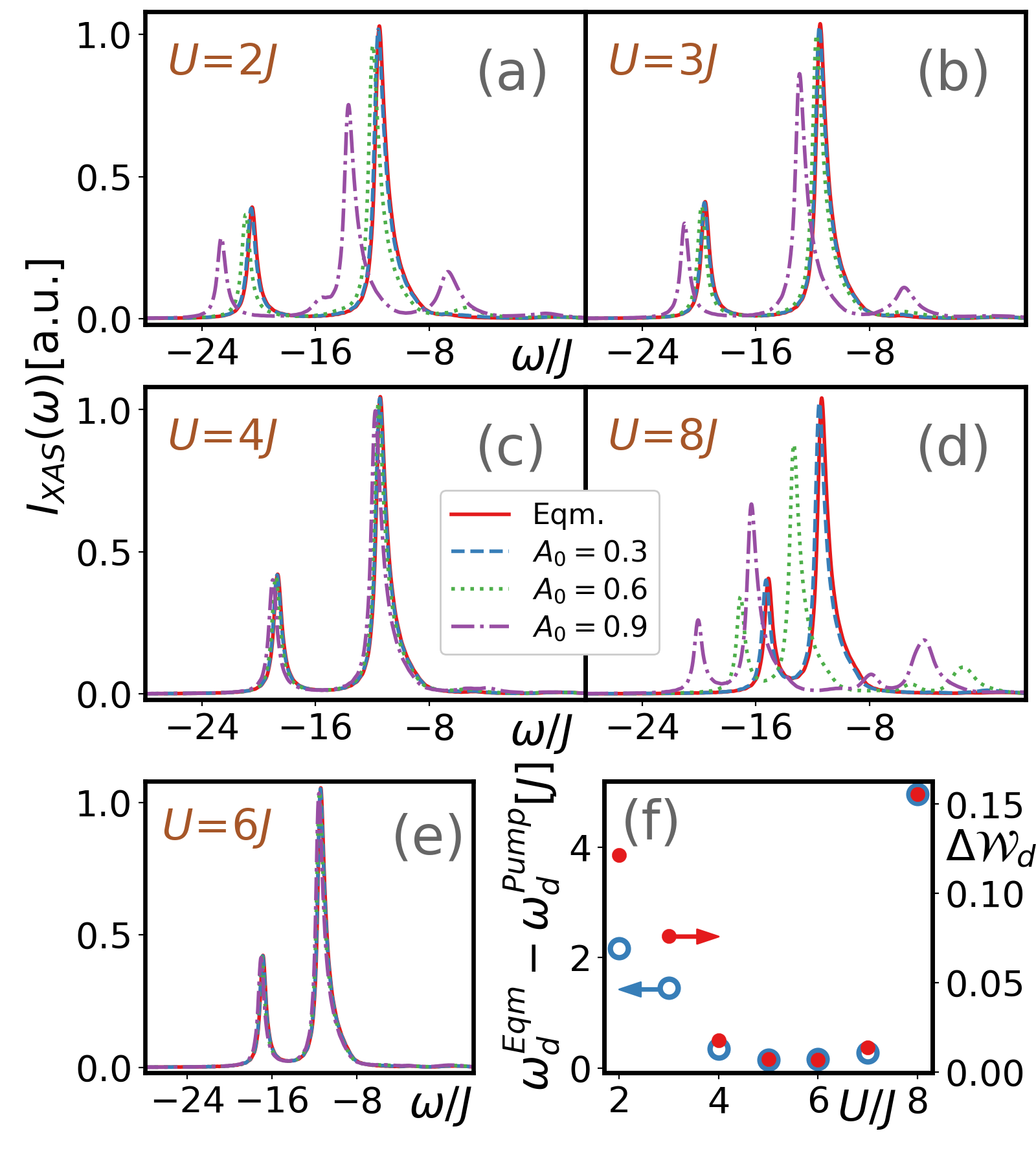}
        \caption{
            (a)-(e) The NE-XAS for systems at $\bar{n}_f\!=\!1/4$ filling under different $A_0$'s where Eqm. stands for static XAS.
            (f) The change of weight (filled symbols) and frequency (empty symbols) of the doubly occupied state with filling $\bar{n}_f\!=\!1/4$ under pump pulse influence $A_0\!=\!0.9$.
            The pump pulse is set to $\Omega t_0\!=\!6$, width $\tau\!=\!3t_0$, and time delay $t_d\!=\!6t_0$.
            The core hole potential is set to $V_\text{ch}\!=\!12J$.
        }
        \label{fig:TR16W6}
    \end{center}
\end{figure}

As the system moves away from half-filling, it is known that the ground state is metallic.
Our calculation of photoemission spectrum also gives the same conclusion as shown in Fig.~\ref{fig:OneThirdDOS}.
Similar to the half filling case, the peaks in the spectrum shift roughly the energy gain from the pump pulse.
The DOS shows no gap near the Fermi energy and a Hubbard satellite with a gap proportional to $U$.
The NE-XAS under various pump pulse fluence is shown in Fig.~\ref{fig:TR16W6} with frequency $\Omega t_0\!=\!6$.
The spectrum is less affected by the pump pulse as the interaction increases from $U\!=\!2J$ to $6J$.
From Fig.~\ref{fig:TR16W6}(b), the results show additional absorption peak emerging around $\omega_{s}\!+\!6J$.
As the interaction gets stronger, the NE-XAS start to respond to the pulse frequency in both change of frequency and the weight of
doubly-occupied bound state, as shown in Fig.~\ref{fig:TR16W6}(f).
The resonance happens around $U\!\sim\!\Omega+2J\!=\!8J$ because the gap between the valance band and Hubbard satellite shown in Fig.~\ref{fig:OneThirdDOS}.
We also expect both the change of frequency and the weight of doubly-occupied bound state to be decreased as the interaction becomes larger than $8J$ which agrees with the result shown in Fig.~4(c) in the main text.

\section{Experimental parameters}
In the article, we use $J$ as the energy unit which is about $J\!\approx\!300meV$ and gives the time unit $t_0\!\approx\!2.2fs$.
The frequency used is around $\hbar\Omega\in[4,6]J\!\approx\![1.2, 1.8]eV$.
The pump pulse we consider have $\tau\!\in\![2t_0,3t_0]\!\approx\![4.4, 6.6]fs$ and the full width at half maximum is
about $2.35\tau\!\approx\![10.34,15.51]fs$ which is smaller than the commonly used experimental values of about $30\!\sim\!50fs$
but is within the range that used in the recent TR experiment on $\mathrm{Y}\mathrm{Ba}_{2}\mathrm{Cu}_{3}\mathrm{O}_{7-\delta}$~\cite{Carbone20161}.
The intensity of the pump pulse can be estimated, for $A_0\in[0.05, 0.5]$ considered here, the pump pulse fluence is around $[9,90]MeV/cm$ for $\Omega t_0\!=\!6$.

The proposed materials in the experiment can be some corner-sharing chains, such as strontium cuprate family $\mathrm{Sr}_2\mathrm{Cu}\mathrm{O}_{3+\delta}$~\cite{kidd2008,kim2006,kim1997,motoyama1996,kim1996}.
Also, there  are  many compounds having the edge-sharing Cu–O chains: for instance, $\mathrm{Li}_2\mathrm{Cu}\mathrm{O}_2$ and $\mathrm{Cu}\mathrm{Ge}\mathrm{O}_3$~\cite{hase1993,keren1993}.
For those materials, the typical nearest neighbor hopping $J$ is around $0.3\!\sim\!0.4eV$ and the Hubbard interaction can be up to $U\!\approx\!2.4\!\sim\!4eV$.
In these type of materials, the separation between two major XAS peaks can be around $4eV$ as the core-valance coupling is estimated around $V_\text{ch}\!\approx\!1.3U$ for an insulating state in equilibrium.
Foe metallic state, the separation will vary according to both $V_\text{ch}$ and $U$ as we show in Fig.~1e of the main text.
Another candidate for the experiments can be polydiacetylene chains~\cite{barcza2010,barcza2013} which is also a Mott insulator despite long range interaction might be presence in those polymers.
Here, the hopping $J$ is around $2.0\!\sim\!2.4eV$ and the Hubbard interaction is roughly $U\!\approx\!5\!\sim\!6eV$ which is in the weak coupling regime.


\bibliography{LANL}